\documentclass[12pt]{article}

\usepackage{amssymb}
\usepackage{graphicx}

\begin{document}

\title{\textbf{Absorption and Direct Processes in Chaotic Wave Scattering}}




\author{R. A. M\'endez-S\'anchez$^{*}$, G. B\'aez$^{\dagger}$, M. Mart\'{\i}nez-Mares$^{**}$ \\ 
{\footnotesize\emph{
$^{*}$Instituto de Ciencias F\'{\i}sicas, Universidad Nacional Aut\'onoma de M\'exico, A. P. 48-3, 62251,}} \\ {\footnotesize\emph{Cuernavaca, Morelos, M\'exico}} \\ 
{\footnotesize\emph{
$^{\dagger}$Departamento de Ciencias B\'asicas, Universidad Aut\'onoma Metropolitana-Azcapotzalco, A. P.}} \\ {\footnotesize\emph{55-534, 09340 M\'exico D. F., Mexico}} \\
{\footnotesize\emph{
$^{**}$Departamento de F\'{\i}sica, Universidad Aut\'onoma
Metropolitana-Iztapalapa, A. P. 55-534, 09340}} \\ {\footnotesize\emph{M\'exico D. F., Mexico}}
}

\date{\empty}

\maketitle

\begin{abstract}
Recent results on the scattering of waves by chaotic systems with losses and direct processes are discussed. We start by showing the results without direct processes nor absorption. We then discuss systems with direct processes and lossy systems separately. Finally the discussion of systems with both direct processes and loses is given. We will see how the regimes of strong and weak absorption are modified by the presence of the direct processes.
\end{abstract}

\vspace{0.5cm}

{\em \noindent We dedicate this paper to Don Leopoldo Garc\'{\i}a-Col\'{\i}n in his 80th birthday.}

\section*{{\normalsize I. INTRODUCTION}}

Since many years, Random Matrix Theory (RMT) has been applied to a plethora of systems in physics such as atomic nuclei, atoms, and molecules~\cite{Brodyetal81,Guhretal98,Metha}. The only assumption is that the underlying ray dynamics is chaotic. More recently it found application in other undulatory systems as mesoscopic transport, 
microwaves, sound and elastic waves, among others~\cite{Stoeckmann,Mendez-Sanchezetal2003,Kuhletal2005,Hul2004,Hul2005,Hemmady2005-1,Hemmady2005-2,KuhlReview,Barthelemy2005-1,Barthelemy2005-2,Fink,Graf}. 
The way in which RMT applies in the different systems, changes from one application to another. To predict the experiment, particular characteristics of each system should be taken in to account as the presence or absence of symmetry, interactions, etc.

RMT yields universal predictions for the fluctuations but these fluctuations appear over an average. 
The average should be then taken into account as a non-universal feature. 

In the case of the energy levels the mean level density is non-universal and depends on each system while the fluctuations are described by the Gaussian ensembles. In the case of scattering systems the average of the $S$-matrix is non-universal but the fluctuations are described by the circular ensembles~\cite{Mello}. These non-universal features should be taken into account to fit experiments. A non-vanishing average of the $S$-matrix yields information of prompt responses in the scattering that can come from an imperfect coupling between the scattering channels and the inner region~\cite{Mendez-Sanchezetal2003,Kuhletal2005}.

As shown in several recent experiments~\cite{Mendez-Sanchezetal2003,Kuhletal2005,Hul2004,Hul2005,Hemmady2005-1,Hemmady2005-2} in mesoscopic systems with classical waves, apart from non-universal features, losses appear. They can be taken into account at least in two different ways. On the one hand, imaginary potentials to simulate absorption (or amplification) can be introduced. On the other hand fictitious (non-measurable) leads can be added to the system. Both approaches are equivalent in certain limit.

When only equilibrated responses are present, the scattering matrix $S_0$ is unitary, its average is zero and 
it is distributed uniformly ($p(\theta)=1/2\pi$). The sub-index indicates the absences of direct processes coming
from prompt responses. Unfortunately this ideal case is not common in the experiment; real systems present direct 
processes coming from prompt responses. Also, in macroscopic systems, losses affect the experimental results. 
In the next section Poisson's kernel will be revisited; in Sec.~III.
the results for systems with 
absorption will be given. In Sec.~IV.
we will give some results of systems in 
which both, losses and direct processes are present. An alternative route was done in 
Refs.~\cite{Hemmady2005-1,Hemmady2005-2} in which the direct processes were extracted from the experiments to 
fit the RMT predictions.

\section*{{\normalsize II. SYSTEMS IMPERFECT COUPLING: THE POISSON'S KERNEL}}

\begin{figure}[h!]
 \includegraphics[width=0.7\columnwidth]{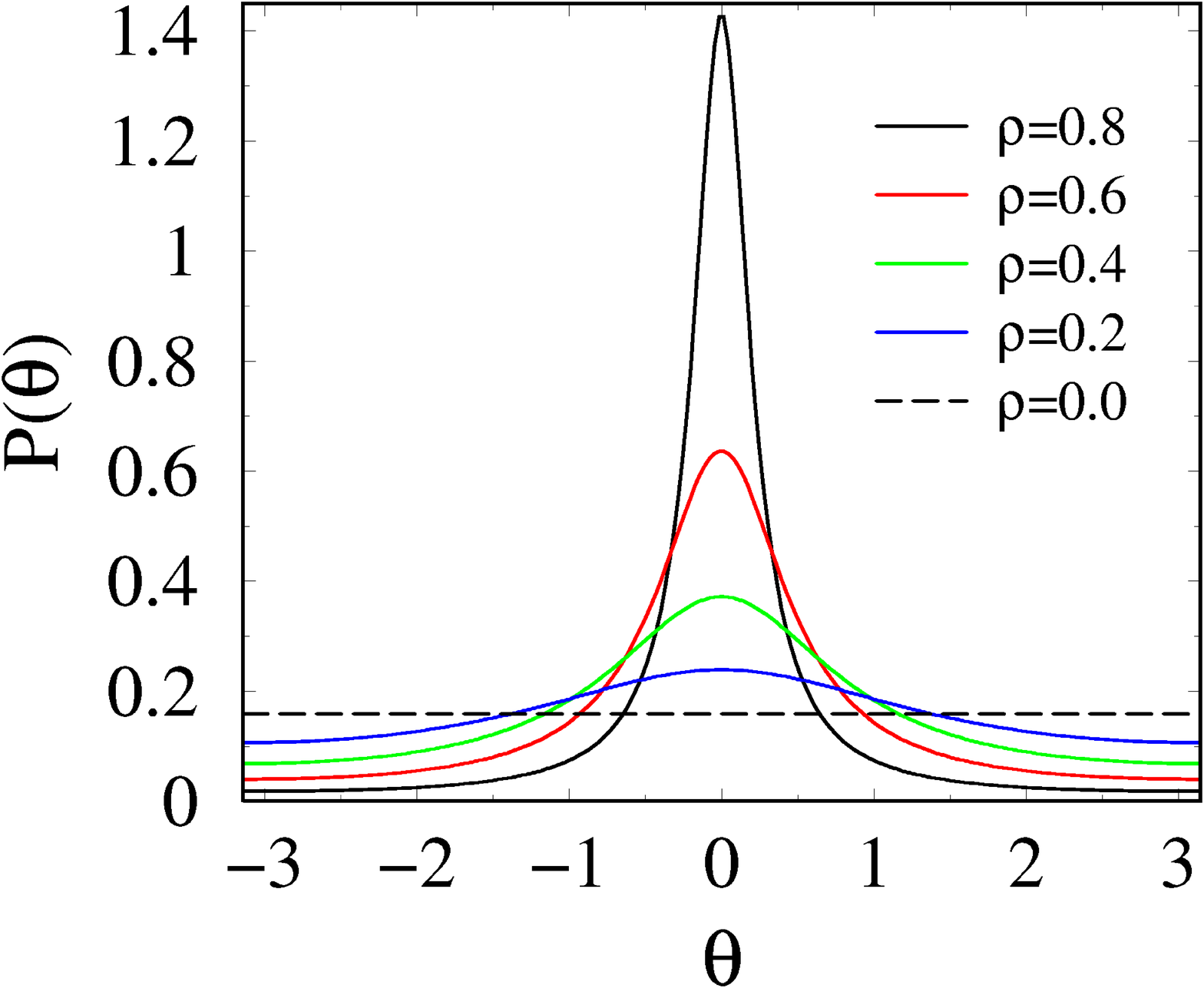}
 \caption{Poisson's kernel for different values of $\rho=|r_p|$. We used $\arg \langle S \rangle =0$.}
 \label{Fig.KernelOfPoisson}
\end{figure}

The statistical distributions of the $S$-matrix with direct processes (or with imperfect coupling) is given by the Poisson's  kernel. In the one-channel, unitary case (without absorption nor amplification) the $S$-matrix is parametrized as $S(E)=e^{i\theta(E)}$, and the Poisson's kernel reads
\begin{equation}
p(\theta)=\frac{1}{2\pi}
\frac{1-\left|\langle S\rangle\right|^2}{\left| S-\langle S \rangle \right|^2},
\label{Eq.PoissonKernel}
\end{equation}
where the brackets refer to ensemble average. A plot of this distribution is given in 
Fig.~\ref{Fig.KernelOfPoisson}. When the system is ergodic, the ensemble averages can be substituted by energy 
averages; in the experiments, however, it is common to perform both averages. 

\begin{figure}[h!]
 \includegraphics[width=0.7\columnwidth]{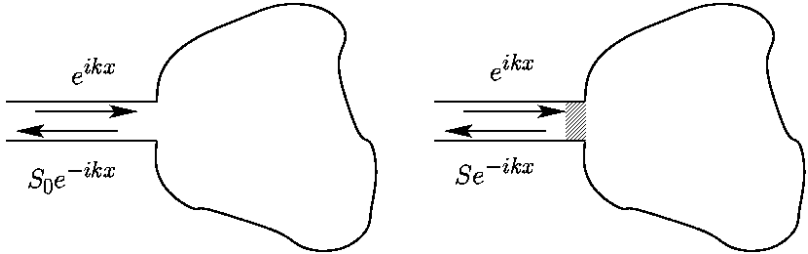}
 \caption{The scattering matrix of the system at the right can be obtained form the scattering matrix of the 
system at the left by composing it with the scattering matrix $S_p$  esoof the barrier. The same scheme is valid when absorption is present; $S$ changes to $\tilde S$ and $S_0$ to $\tilde S_0$.}
 \label{Fig.TransformationWithoutLosses}
\end{figure}

The Poisson's kernel can be obtained using different approaches. One of them is calculating the Jacobian of the 
transformation between the $S$-matrices of a system without direct processes and another with direct processes. 
This approach has a physical interpretation shown in 
Fig.~\ref{Fig.TransformationWithoutLosses}: the $S$-matrix of the system with direct processes can be obtained 
from the composition of the $S_0$-matrix of the system without losses and the $2\times2$ scattering matrix $S_p$ of a barrier
\begin{equation}
 S_p=\left( \begin{array}{ll}
             r_p & t'_p \\ 
             t_p & r'_p      
            \end{array}
     \right).
\label{Eq.S_p}
\end{equation}
The composition is given by
\begin{equation}
 S=r_p+t'_p\frac{1}{1-S_0 r'_p} S_0 t_p.
 \label{Eq.Transformation1Ch}
\end{equation}
One can notice that $\langle S \rangle =r_p$ enters into Eq.~(\ref{Eq.PoissonKernel}). When $\langle S \rangle =0$ one gets that $S=S_0$ and then $S$-matrix is distributed as $S_0$, i.e. uniformly.

\begin{figure}[h!]
 \includegraphics[width=0.7\columnwidth]{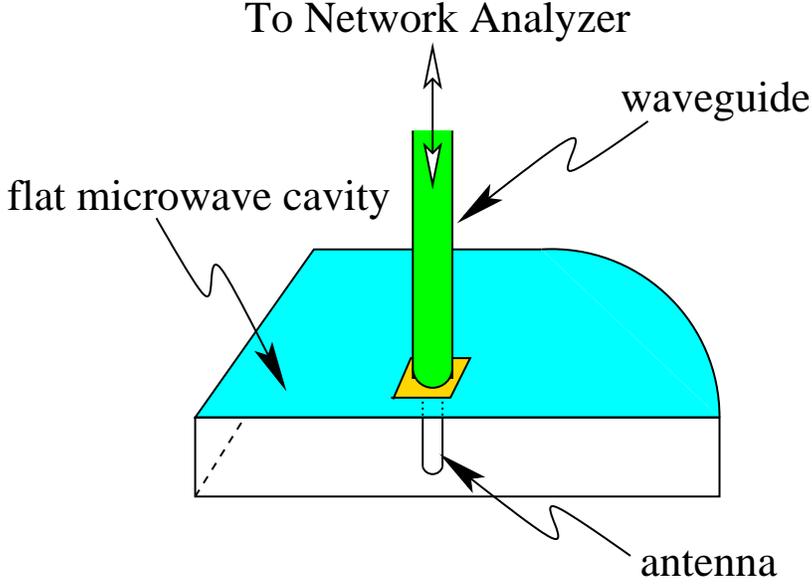}
 \caption{Experimental realization of Eq.~(\ref{Eq.Transformation1Ch}). The coupling between the antenna and the microwave cavity yields the direct processes; since the flat microwave cavity has the shape of the stadium, the distribution of its scattering matrix is chaotic.
}
 \label{Fig.ExperimentalSetup}
\end{figure}

In several experiments, as the shown in Fig.~\ref{Fig.ExperimentalSetup} with flat microwave cavities, the direct processes come from direct reflections at the antenna but other sources of direct processes cannot be disregarded. Thus the coupling $T_a$ between the antenna and the cavity can be modeled by the $S_p$-matrix of Eq.~(\ref{Eq.S_p}) as $T_a=1-|\langle S\rangle|^2$.

The results presented above can be generalized for $N$ channels. A similar transformation to that of Eq.~(\ref{Eq.Transformation1Ch}) is valid in the case of $N$ channels. The distribution of the $n\times n$ scattering matrix with direct processes is given by
\begin{equation}
\label{Eq.dP}
{\rm d} P^{(\beta )}_{\langle S\rangle}(S)
=p^{(\beta )}_{\langle S\rangle}(S)\, {\rm d} \mu _{\beta }(S) ,
\end{equation}
where 
\begin{equation}
\label{Eq.PoissonNChannels}
p^{(\beta)}_{\left\langle S\right\rangle }(S)
= \frac
{[{\rm det}(I_n-\left\langle S\right\rangle \left\langle S\right\rangle ^{\dag})]^{(\beta N+2-\beta)/2}}
{|{\rm det}(I_n-S\left\langle S\right\rangle ^{\dag})|^{(\beta N+2-\beta)}},
\end{equation}
is the Poisson's kernel for $N$ channels and ${\rm d} \mu _{\beta }(S)$ is the invariant measure of the circular ensembles~\cite{Dyson}. 

\section*{{\normalsize III. SYSTEMS WITH LOSSES AND PERFECT COUPLING}}
\label{S1}
    
When absorption is present the scattering matrix is non-unitary. 
In the one-channel case, $N=1$, the scattering matrix can be parametrized as
\begin{equation}
 \tilde S_0=\sqrt{R_0}e^{i\theta}
\end{equation}
where $R_0$ is the reflection coefficient and $\theta$ is twice the phase shift. The tilde is used to indicate that the $S$-matrix is non-unitary and, as before, the sub-index indicate that the direct processes vanish i.e., $\langle \tilde S \rangle=0$. The distribution of the scattering matrix in this case is 
\begin{equation}
dP^{(\beta)}_0({\tilde S}_0)=p^{(\beta)}_0(R_0)\, 
dR_0\,\frac{d\theta_0}{2\pi},
\label{eq:dP0}
\end{equation}
and the interest is on the distribution $p^{(\beta)}_0(R_0)$ of the reflection coefficient $R_0=\tilde S_0 \tilde S_0^\dagger$. The strength of the losses is quantified by $\gamma$ that, at the moment, will be taken as a fitting parameter. 
When the underling classical dynamics is chaotic, $R_0$ fluctuates according to
\begin{equation}
\label{eq:p0(R0)-b12}
p^{(\beta)}_0(R_0) = \frac 2{ (1-R_0)^2 }
P^{(\beta)}_0\left(\frac{1+R_0}{1-R_0}\right); \qquad W_{\beta}(x)= \int_x^{\infty}dx\, P^{(\beta)}_0(x),
\end{equation}
were $\beta=1,2$ indicates the presence or absence of time reversal symmetry and~\cite{SavinSommers2003,BaezMartinez-MaresMendez-Sanchez}
\begin{eqnarray}
\label{eq:W(x)-2}
W_1(x) & = & \frac{x+1}{4\pi}\Big[
f_1(w)g_2(w)+f_2(w)g_1(w) \nonumber \\ 
& + & 
h_1(w)j_2(w)+h_2(w)j_1(w)
\Big]_{w=(x-1)/2},\quad \\
W_2(x) &=& 
\frac 12 
e^{-\gamma x/2}
\left[ e^{\gamma/2}(x+1)-e^{-\gamma/2} (x-1) \right]
\nonumber
\end{eqnarray}
with $\qquad f_q(w) = \int_{l_q}^{u_q} dt \frac{\sqrt{t|t-w|}e^{-\gamma t/2}\left[1-e^{-\gamma} + t^{-1}\right]}{(1+t)^{3/2}}
$, $\qquad g_q(w) = \int_{l_q}^{u_q} dt \frac 1{\sqrt{t|t-w|}}
\frac{e^{-\gamma t/2}}{(1+t)^{3/2}}$, \\ $h_q(w)  = \int_{l_q}^{u_q} dt
\frac{\sqrt{|t-w|}e^{-\gamma t/2}[\gamma + (1-e^{-\gamma})(\gamma t - 2)]}{\sqrt{t(1+t)}} $,\,\, and\,\, $j_q(w) = \int_{l_q}^{u_q} dt \frac 1{\sqrt{t|t-w|}} \frac{e^{-\gamma t/2}}{\sqrt{1+t}}$.
Here $q=1,\,2$, $l_1=w$, $l_2=0$, $u_1=\infty$, and $u_2=w$. This formalism agrees with that obtained previously by Brouwer and Beenakker for $\beta=2$~\cite{BeenakkerBrouwer2001}. In the case of two channels, $N=2$, the distribution of the reflection, $p(R_1,R_2)$ becomes more complicated; it was found by Beenaker and Brouwer~\cite{BeenakkerBrouwer2001,BrouwerBeenakker1997}. The distribution of the reflection eigenvalues was also found for any number of channels for $\beta=2$~\cite{SavinSommers} and for the strong absorption limit, any number of channels for $\beta=1,2$ by Mello and coworkers~\cite{KoganMelloLiqun}.

\section*{{\normalsize IV. SYSTEMS WITH BOTH LOSSES AND DIRECT PROCESSES}}
\label{S2}

When the system is non-unitary and has direct processes the way to proceed is similar to that when the scattering matrix is unitary. A non-unitary matrix $\tilde S_0$ without direct processes is composed with the unitary $S_p$-matrix of the direct processes to get a scattering matrix with both direct processes and losses. A similar transformation as Eq.~(\ref{Eq.Transformation1Ch}) is valid for non-unitary matrices but the inverse transformation
\begin{equation}
 \tilde S_0(\tilde S) = \frac{\tilde S-\langle \tilde S\rangle }
{1-\langle \tilde S\rangle \tilde S},
\label{Eq.TransformationLosses}
\end{equation}
is needed. The scattering matrix with both losses and direct processes can be parametrized as
\begin{equation}
 \tilde S=\sqrt{R}e^{i\theta},
\end{equation}
and its distribution is given by 
\begin{equation}
dP^{(\beta)}_{\langle {\tilde S}\rangle}({\tilde S}) = 
p^{(\beta)}_{\langle {\tilde S}\rangle}({\tilde S})\, dR\, \frac{d\theta}{2\pi},
\label{eq:dP}
\end{equation}
where
\begin{equation}
 p^{(\beta)}_{\langle {\tilde S}\rangle} (\tilde S)=\frac{1}{2 \pi}\left(\frac{1-\left|\langle \tilde S\rangle\right|^2}{\left| \tilde S-\langle \tilde S \rangle \right|^2}\right)^2 p_0(R_0(R))
\end{equation}
and $R_0(R)$ should be determined from Eq.~(\ref{Eq.TransformationLosses}). Some plots of the last equation for different absorption and coupling parameters is given in Ref.~\cite{Kuhletal2005} among with a comparison with numerical and experimental results using chaotic flat microwave cavities. The marginal distributions of the last equation obtained with numerical integration are also given there.

\begin{figure}[h!]
\includegraphics[width=0.7\columnwidth]{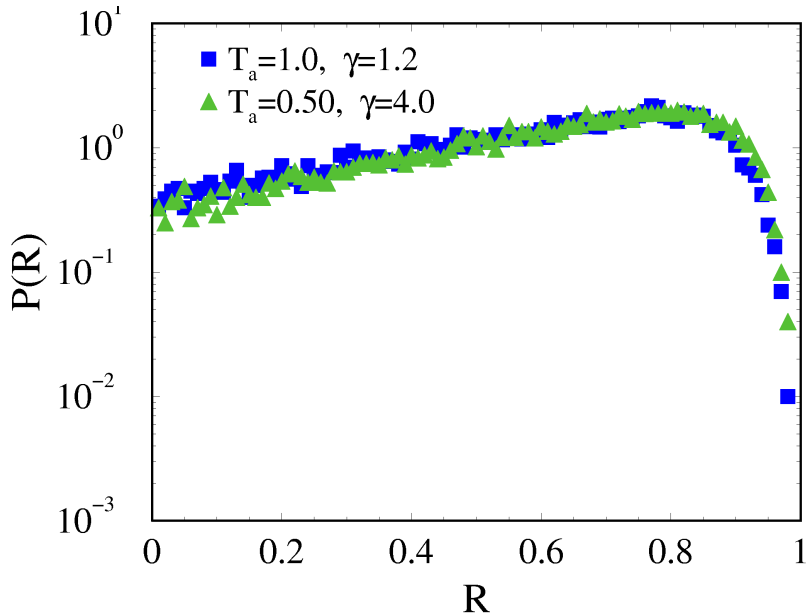}
\caption{Distribution of the reflection coefficient, one with and another without direct processes. The case of $\langle \tilde S_0\rangle=0$ and $\gamma=1.2$ is given by the squares; the triangles correspond to $\langle \tilde S_0\rangle=0.5$ and $\gamma=4$.
}
\label{Fig.BeCarefull}
\end{figure}

Now we will show that different parameters of the direct processes and absorption could give similar results. In Fig.~\ref{Fig.BeCarefull} we show numerical results for the distribution of the reflection coefficient for a system without direct processes, $\langle \tilde S\rangle=0$, and weak absorption, $\gamma=1.2$ and another with some direct processes, $\langle \tilde S\rangle=0.5$ and with stronger absorption, $\gamma=4$. Those results were obtained using the {\em Heidelberg approach} with one open channel and several fictitious leads that simulate the losses~\cite{Caio}. As can be seen in this figure, the distribution of the reflection coefficient is quite similar for both cases. Since the direct processes appear in all transport systems (classical or quantum), this case shows that it is absolutely necessary to know quantify the direct processes are in each experimental system since transport properties, like the conductance, can be interpreted erroneously when the direct processes are not taken into account.

In the one-channel case there are two points missing, one related to the evaluation of the parameter $\gamma$ and another related to the regimes defined by (a) the losses and (b) the direct processes. The first question was addressed using two different approaches, one by adjusting the fidelity~\cite{Seligman} and another using the average of the reflection coefficient~\cite{BaezMartinez-MaresMendez-Sanchez}. What is needed to quantify the strength of the absorption is a measurable parameter with a correspondence one to one with $\gamma$. One possibility relies on the average of the reflection coefficient $\langle R \rangle$ and another in the fidelity. The second question was addressed only in Ref~\cite{BaezMartinez-MaresMendez-Sanchez}; the direct processes introduce in the game a new time scale and redefine the regimes of weak and strong absorption are defined as $\gamma<<T_a$ and $\gamma>>T_a$, respectively.

In the case of $N$-channels, the Poisson's kernel can be generalized in the same way for non-unitary matrices. The result was obtained as the Jacobian of the transformation that connects a non-unitary scattering matrix without direct processes and a non-unitary matrix with direct processes that are connected by $\langle \tilde S \rangle$. 

\section*{{\normalsize V. CONCLUSIONS}}

We discussed the recent results on the scattering waves in presence of both, losses and direct processes. 
The distribution of the non unitary scattering matrix $\tilde S$ with direct processes is given by the square of the Poisson's kernel. It is completely characterized by the average $\langle \tilde S \rangle$ and by the strength of the losses $\gamma$. The results presented here show that the direct processes cannot be avoided in the analysis of transport properties like the conductance since similar results could be obtained for systems with and without direct processes but for different parameters of the losses. Also, the regimes of strong and weak absorption change when direct processes are present.


\section*{{\normalsize ACKNOWLEDGMENTS}}

This work was supported by DGAPA-UNAM under projects IN111308 and by CONACyT under project 79613.


\end{document}